\documentclass[aps,prx,amsmath,twocolumn,groupedaddress,letterpaper,floatfix]{revtex4-1}
\usepackage{graphicx,color}
\usepackage{verbatim}
\usepackage{amssymb}   % for math
\usepackage{amsmath}
\usepackage{amsfonts}
\usepackage{mathdots}
\usepackage{hyperref}
\usepackage{xcolor}
\usepackage{epsfig}
\usepackage{braket}
\usepackage{bm}

\begin{document}
\title{Valley-polarized Josephson Junctions as gate-tunable $0$-$\pi$ qubit platforms}
\author{Zhong-Chang-Fei Li}\thanks{These authors contributed equally to this work.}
\author{Yu-Xuan Deng}\thanks{These authors contributed equally to this work.}
\author{Zi-Ting Sun}
\author{Jin-Xin Hu}\thanks{jhuaw@connect.ust.hk}
\author{K. T. Law}\thanks{phlaw@ust.hk}
\affiliation{Department of Physics, Hong Kong University of Science and Technology,
Clear Water Bay, Hong Kong SAR, China}
\date{\today}

\begin{abstract}
Recently, gate-defined Josephson junctions based on magic-angle twisted bilayer graphene (MATBG) have been fabricated. In such a junction, local electrostatic gating can create two superconducting regions connected by an interaction-driven valley-polarized state as the weak link. Due to the spontaneous time-reversal and inversion symmetry breaking of the valley-polarized state, novel phenomena such as the Josephson diode effect have been observed without applying external fields. Importantly, when the so-called nonreciprocity efficiency (which measures the sign and strength of the Josephson effect) changes sign, the energy-phase relation of the junction is approximate $F(\phi) \approx \cos(2\phi)$ where $F$ is the free energy and $\phi$ is the phase difference of the two superconductors. In this work, we show that such a MATBG-based Josephson junction, when shunted by a capacitor, can be used to realize the long-sought-after $0$-$\pi$ qubits which are protected from local perturbation-induced decoherence. Interestingly, by changing the junction parameters, transmon-like qubits with large anharmonicity can also be realized. In short, by utilizing the novel interaction-driven valley-polarized state in MATBG, a single gate-defined Josephson junction can be used to replace complicated superconducting circuits for realizing qubits that are protected from local perturbations. 
\end{abstract}
\maketitle

\emph{Introduction.}---Superconducting circuits provide versatile platforms for exploring quantum physics \cite{you2011atomic,houck2012chip, PhysRevLett.122.210401} and quantum information processing \cite{RevModPhys.73.357,kimble2008quantum,gambetta2017building,krantz2019quantum,doi:10.1126/science.abb9811,doi:10.1146/annurev-conmatphys-031119-050605}. In these circuits, two energy levels, such as the ground state and the first excited state, can serve as qubit states for encoding quantum information, provided that the two states can be reliably distinguished and selectively controlled \cite{https://doi.org/10.1002/cta.2359,PRXQuantum.2.040204}. The crucial components in these circuits are Josephson junctions \cite{RevModPhys.46.251}, which can introduce nonlinearity in the energy levels without energy dissipation \cite{devoret2004superconducting}.

%%%%%%%%%%%%%%%%%%%%%
\begin{figure}[ht]
\centering \includegraphics[width= 1\linewidth]{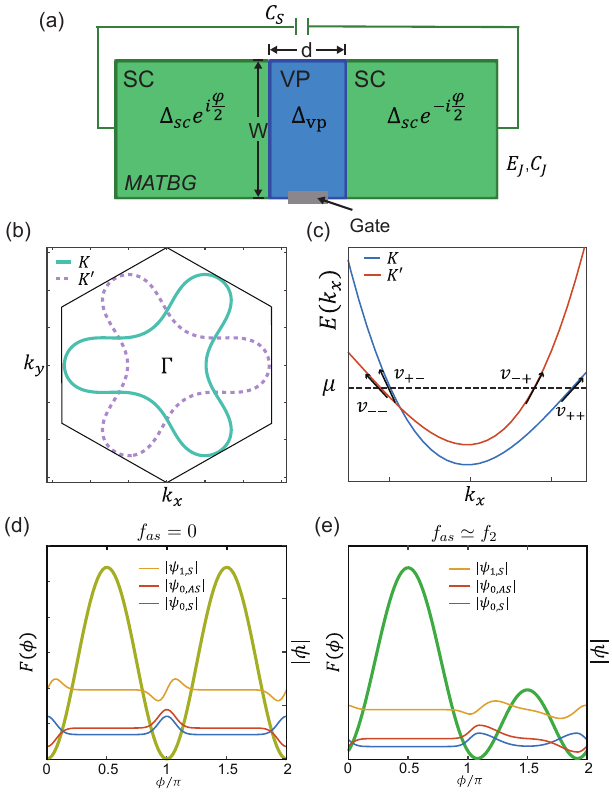} 
\caption{(a) A schematic picture of valley-polarized Josephson qubit. For the Josephson junction, the central weak link is partially valley-polarized (VP) with valley polarization order parameter $\Delta_{\mathrm{vp}}$, while the left and right regions are conventional s-wave superconductors (SC) with pairing potential $\Delta_{\mathrm{sc}}$. (b) The trigonal warping Fermi surface around $K$ and $K'$ valleys of MATBG at half filling. (c) Band dispersion of the $k_y=0$ channel for the valley-polarized weak-link region, with the black arrows indicating the Fermi velocity $v_{\tau,\alpha}$. Here, $\tau = \pm$ denotes the $\pm K$ valleys and $\alpha=\pm$ labels the left and right movers of electrons, respectively. (d)-(e) The schematic energy-phase relations and eigenstates of valley-polarized Josephson qubit for two different parameter regimes. The left y-axis is the Josephson free energy $F(\phi)$, and $\psi_{0,S}$, $\psi_{0,AS}$ and $\psi_{1,S}$ denote the symmetric and antisymmetric qubit states and the third energy level, respectively. The energy-phase relations in (d) and (e) resemble those of $0$-$\pi$ qubits and flux qubits respectively. }
\label{fig:fig1} 
\end{figure}
% 
%%%%%%%%%%%%%%%%%%%%%%%%%%%%%%%%%%

Currently, there are several promising qubit platforms with their own strengths and limitations. For example, a charge qubit \cite{bouchiat1998quantum,nakamura1999coherent,RevModPhys.93.025005} or a transmon qubit \cite{PhysRevA.76.042319,PhysRevB.77.180502,RevModPhys.93.025005} can be constructed using a single Josephson junction shunted by a capacitor. In the case of a charge qubit, the shunted capacitor is small so that it is sensitive to charge noise but has large anharmonic energy levels, while the shunted capacitor for a transmon is large so that a transmon is robust against charge noise at the expense of a smaller anharmonicity of energy levels. On the other hand, a flux qubit 
\cite{mooij1999josephson,PhysRevB.60.15398,PhysRevB.75.140515,yan2016flux} is built by a superconducting loop which contains a series of Josephson junctions and penetrated by external magnetic flux. The energy levels are charge-insensitive and have large anharmonic. However,  fluctuations of the magnetic flux introduces additional decoherence. To overcome decoherence, $0$-$\pi$ qubits with two degenerate ground states \cite{kitaev2006protected,PhysRevA.87.052306,PhysRevB.90.094518,groszkowski2018coherence} as the qubit states were proposed, in which the qubits are protected from local-noise decoherence as pointed out by Kitaev \cite{kitaev2006protected}. However, even with some experimental success, the realization of $0$-$\pi$ qubits is hindered by the complexity of the circuit design \cite{PRXQuantum.2.010339}. 
%By using the following novel valley polarized Josephson junction, we can realize $0$-$\pi$ qubit by a single Josephson junction without external magnetic field and design a transmon-like qubit with large anharmonicity without requiring unconventional properties of superconductors.

In this work, we point out that recent experimental advancements in fabricating gate-defined Josephson junctions based on magic-angle twisted bilayer graphene (MATBG) have opened a novel pathway for engineering highly gate-tunable qubits. In such a MATBG-based Josephson junction \cite{2023NatCo..14.2396D}, electrostatic gating can define two superconducting regions which are connected by a valley-polarized, time-reversal symmetry breaking, Josephson junctions through electrostatic gating without applying external magnetic fields \cite{2023NatCo..14.2396D}. Moreover, the valley-polarization lifts the valley degeneracy spontaneously \cite{PhysRevX.8.031089} and gives rise to unconventional phenomena such as the generation of the $\phi_0$-Josephson junction \cite{Yingming_phi_0_junction} and the Josephson diode effect \cite{Jinxin_JJ_Diode}. For the Josephson diode effect, the critical (or the maximum) supercurrent flowing in the positive direction $I_{c+}$ differs from the critical supercurrent $I_{c-}$ flowing in the negative direction. The nonreciprocity efficiency $\eta = (I_{c+} - |I_{c-}|)/(I_{c+} + |I_{c-}|)$ can be as high as 30\% in the experiment \cite{2023NatCo..14.2396D, Jinxin_JJ_Diode}.

To explain the Josephson diode effect in the gate-defined Josephson junction of MATBG \cite{2023NatCo..14.2396D}, we showed previously that the unconventional energy-phase relation $F(\phi)$ of the Josephson junction has the form \cite{Jinxin_JJ_Diode}
\begin{equation}
\label{free energy}
F(\phi)=f_{\mathrm{vp}}\cos\phi+f_{\mathrm{as}}\sin\phi+f_2 \cos 2\phi.
\end{equation}
Here, $f_{\mathrm{vp}}$ denotes the first harmonic term that can be controlled by valley polarization through gating,  $f_{\mathrm{as}}$ is the asymmetric term from intra-valley inversion symmetry breaking which is also gate-tunable, whereas $f_2$ is the second harmonic term. Importantly, the parameters of the Josephson energy can be gate-tuned to be dominated by the $\cos 2\phi$ term when $f_{\mathrm{vp}}=f_{\mathrm{as}}=0$. This happens when the nonreciprocal efficiency changes sign  \cite{Jinxin_JJ_Diode}. In this work, we demonstrate that when the junction is shunted by a capacitor,  a $0$-$\pi$ qubit is obtained as schematically shown in [Fig.~\ref{fig:fig1}~(a)]. The energy phase relation of such a qubit is schematically shown in [Fig.~\ref{fig:fig1}~(d)]. The advantage of the setup in Fig.~\ref{fig:fig1}~(a) is that a $0$-$\pi$ qubit can be realized by a single Josephson junction without sophisticated circuit design \cite{kitaev2006protected,PhysRevA.87.052306,PhysRevB.90.094518,groszkowski2018coherence}. This is possible because of the realization of the novel interaction-driven, valley-polarized state of matter in MATBG \cite{Yingming_phi_0_junction, Jinxin_JJ_Diode}.  Moreover, the time-reversal symmetry is spontaneously broken by the interaction-driven valley polarization rather than an external magnetic field. This can reduce the decoherence caused by magnetic flux fluctuations. 

On the other hand, if $f_{\mathrm{vp}}=0$ and $f_{\mathrm{as}}\simeq f_2$, then $F(\phi)$ also have two nearly degeneracy local minima. The energy-phase relation of this regime is shown in  Fig.~\ref{fig:fig1}~(e) which we call the transmon regime. As we show later, the qubit states can have a large energy separation from the higher energy states such that a large anharmonicity can be realized for the transmon regime. Moreover, for both of the two proposals, only conventional s-wave superconductors are required for the superconducting regions, so that the relaxation caused by quasiparticles is exponentially suppressed by the superconducting gap \cite{PhysRevB.84.064517} when the qubit works at temperature $k_BT\ll\Delta_{\mathrm{sc}}$, where $\Delta_{\mathrm{sc}}$ is the superconducting gap. In the following sections, we show how the energy-phase relations, as schematically shown in Fig.~\ref{fig:fig1}~(d) and Fig.~\ref{fig:fig1}~(e), can be realized by gated-defined MATBG Josephson junctions with valley-polarized state as the weak link. We further show how to realize the $0$-$\pi$ qubits and the transmon qubits with large anharmonicity using gate-defined MATBG.

\emph{Model of gate-defined Josephson junction in MATBG.}---We consider a Josephson junction consisting of two superconducting regions at $x<0$ and $x>d$, respectively. The two superconducting regions are connected by a weak link with valley degrees of freedom (at $0<x<d$) as schematically shown in [Fig.~\ref{fig:fig1}(a)]. For simplicity, the superconductors are assumed to have s-wave pairing. The superconducting pairing order parameter can be written as $\Delta_{\mathrm{sc}}(x)=\Delta_{\mathrm{sc}}[e^{i\frac{\phi}{2}}\Theta(-x)+e^{-i\frac{\phi}{2}}\Theta(x-d)]$ which pair electrons from opposite valleys and opposite spins. Furthermore, we suppose that the weak link is in an interaction-driven valley polarized state and the intravalley inversion symmetry can be broken by the trigonal warping of the Fermi surface [as schematically shown in Fig.~\ref{fig:fig1}(b)]. The Hamiltonian  \cite{Jinxin_JJ_Diode} of the weak link at each valley can be expressed as
\begin{equation}
\label{0}
    H_{0,\tau}(\mathbf{k})=\lambda_{0}T_\mathbf{k}+\lambda_{1} A_{\mathbf{k},\tau}+\Delta_{\mathrm{vp}}\tau-\mu,
\end{equation}
 where $\tau=+/-$ labels valley, $T_\mathbf{k}$ is the isotropic part of kinetic energy, $A_{\mathbf{k},\tau}$ is the term that breaks intravalley inversion symmetry, meaning that $A_{\mathbf{k},\tau}\neq A_{\mathbf{-k},\tau}$, and  $\Delta_{\mathrm{vp}}$ represents the valley polarization which lifts the valley degeneracy and breaks time-reversal symmetry. For MATBG, $T_\mathbf{k}=k_x^2+k_y^2$ and $A_{\mathbf{k},\tau}=\tau k_x(k_x^2-3k_y^2)$ denotes the trigonal warping term. The Fermi velocities of the electrons in the valley-polarized weak link are expressed as $v_{\mathrm{vp},\tau\alpha}$, where $\alpha=+/-$ labels the direction of the Fermi velocity. Due to $A_{\mathbf{k},\tau}$, the left moving electrons and right moving electrons within the same valley have different Fermi velocities $v_{\mathrm{vp},\tau\alpha}\neq v_{\mathrm{vp},\tau-\alpha}$ [Fig.~\ref{fig:fig1}~(c)]. The weak link is assumed to be partially valley-polarized as shown in Fig.~\ref{fig:fig1}~(c) which means that the energy degeneracy of the $K$ and $-K$ valleys are lifted and the bands from both valleys are partially filled. The spin degrees of freedom are also assumed to be degenerate. In the following, we assume $k_y=0$ for the analytical illustration of the energy-phase relation. The energy-phase relation for a two-dimensional junction is calculated numerically in a later part of this work.  % The real platform is expected to be a 2D system to have interaction-driven valley polarization, but for illustration we use $k_y=0$ channel to show the influence of $\Delta_{\mathrm{vp}}$ and $A_{\mathbf{k}}$ on the energy phase relation. In latter section, we use 2D TBG as a concrete example. 

For each valley, near the Fermi energy, the linearized electron Hamiltonian for $k_y=0$ is $H_{0,\tau\alpha}(x)=-i\alpha\hbar v_{f,\tau\alpha}(x)\partial_x+\Delta_{\mathrm{vp}}(x)\tau$, where $v_{f,\tau\alpha}(x)=v_{\mathrm{vp},\tau\alpha}(x)\Theta(x)\Theta(d-x)+v_{\mathrm{sc},\tau\alpha}(x)(\Theta(-x)+\Theta(x-d))$, including the Fermi velocities in the valley-polarized state  and the superconducting state, and $\Delta_{\mathrm{vp}}(x)=\Delta_{\mathrm{vp}}\Theta(x)\Theta(d-x)$ is the valley polarization order parameter. To simplify the following discussion, we define two energy scales: $E_T=\frac{\hbar v_f}{d}$ is the Thouless energy and $E_A=\frac{\hbar\delta v_f}{d}$ is the energy scale describing the intravalley inversion symmetry breaking, where $v_f^{-1}=\frac{1}{4}\sum_{\tau\alpha}v_{\mathrm{vp},\tau\alpha}^{-1}, \delta v_f^{-1}=\frac{1}{2}(v_{\mathrm{vp},++}^{-1}-v_{\mathrm{vp},+-}^{-1}+v_{\mathrm{vp},--}^{-1}-v_{\mathrm{vp},-+}^{-1})$. In the Nambu basis $(\psi_{\tau\alpha}(x), \psi_{-\tau-\alpha}^\dagger(x))$, the linearized BdG Hamiltonian of this Josephson junction is 
\begin{equation}
    \label{1}
    H_{\tau\alpha}(x)=\begin{pmatrix}
        H_{0,\tau\alpha}(x) & \Delta_{\mathrm{sc}}(x)\\
        \Delta_{\mathrm{sc}}^{*}(x) & -H_{0,-\tau-\alpha}^{*}(x)
    \end{pmatrix},
\end{equation}
from which the energy-phase relation can be calculated analytically using the scattering matrix method \cite{beenakker1992three}. The outline of the calculations is given below and the details can be found in \cite{Supplementary}.

%%%%%%%%%%%%%%%%%%%%%%%%%%%%%%%%%%%%%%%%%%%%%%%%

\begin{figure}[t]
% \centering 
\includegraphics[width=1\linewidth]{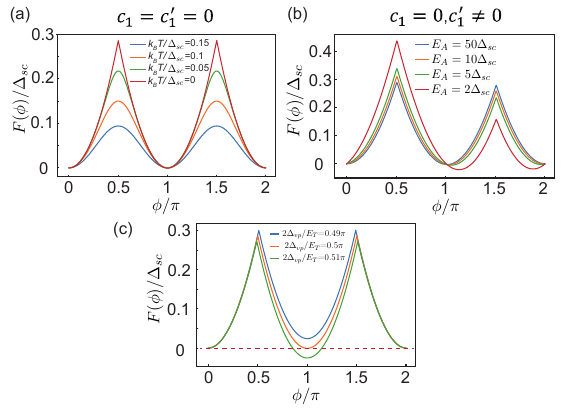}
    \caption{ $\Delta_\mathrm{vp}=2\Delta_\mathrm{sc}$ (a) energy-phase relations at different temperatures with $E_A\gg\Delta_\mathrm{sc}$, $\frac{2\Delta_\mathrm{vp}}{E_T}=\frac{\pi}{2}$. (b) energy-phase relations for different $E_A$ under zero temperature approximation with $\frac{2\Delta_\mathrm{vp}}{E_T}=\frac{\pi}{2}$ (c) The influence of valley polarization on the energy-phase relation. As the valley polarization deviates from $\frac{2\Delta_\mathrm{vp}}{E_T}=\frac{\pi}{2}$, the degeneracy of the two local minima of the free energy is lifted. The results are obtained under zero temperature approximation with $E_A\gg\Delta_\mathrm{sc}$}
    \label{Fig.1D}
\end{figure}

%%%%%%%%%%%%%%%%%%%%%%%%%%%%%%%%%%%%%%%%%%%%%%%%%
\emph{Energy-Phase Relation.}---Based on the scattering matrix method, we illustrate that both $\Delta_{\mathrm{vp}}$ and $A_{\mathbf{k},\tau}$ make the energy-phase relation unconventional. The normal scattering states in the weak-link region can be depicted by the relation $\vec{c}_N^{\text{ out}}=\mathcal{S}_N\vec{c}_N^{\text{ in}}$, where $\vec{c}_N^{\text{ in(out)}}$ describes the incoming (outgoing) states and $\mathcal{S}_N$ is the scattering matrix. There is also Andreev reflection processes $\vec{c}_N^{\text{ in}}=\mathcal{S}_A\vec{c}_N^{\text{ out}}$ at the interface between the weak-link and superconducting region. The details are given in the Supplementary Material. Then the energy-phase relation can be evaluated as \cite{ FURUSAKI1994214,brouwer1997anomalous}
\begin{equation}
    \label{2}
    F(\phi)=-k_B T\sum_{n=0}^{\infty}\ln \det[1-\mathcal{S}_A(i\omega_n,\phi)\mathcal{S}_N(i\omega_n,\phi)],
\end{equation}
where $\omega_n$ are Matsubara frequencies and the Josephson current can be calculated by $I(\phi)=\frac{2e}{\hbar}\frac{\partial F}{\partial \phi}$. 

At finite temperature, for example, at $k_BT=0.15\Delta_{\mathrm{sc}}$, the $\omega_0$ term of the Matsubara sum  dominates in the free energy (Eq.\ref{2}). The corresponding Josephson current can be expressed as $I(\phi)=\frac{2e}{\hbar}\Delta_{\mathrm{sc}}\big( c_1\sin \phi+c'_1\cos \phi+c_2\sin 2\phi \big)$ \cite{Jinxin_JJ_Diode,Supplementary}. Here, the $\sin(\phi)$-term is the conventional Josephson current which is originated from the tunneling of one Cooper pair at each tunneling event. The $\cos(\phi)$-term is the anomalous Josephson current caused by the fact that there is a finite phase difference between the two superconductors at the ground state \cite{Yingming_phi_0_junction}. The $\sin(2\phi)$-term is the second harmonic Josephson current originated from the tunneling of two pairs of Cooper pairs at each tunneling event. The corresponding energy-phase relation $F(\phi)$ has the form of Eq.~\ref{free energy}. With the analytical form of the scattering matrices (See Supplementary Materials), we find that $c_1\propto\cos(\frac{2\Delta_{\mathrm{vp}}}{E_T})\cosh(\frac{\omega_0}{E_A})$ and $c'_1\propto\sinh(\frac{\omega_0}{E_A})\sin(\frac{2\Delta_{\mathrm{vp}}}{E_T})$. At $2\Delta_{\mathrm{vp}}/E_T=\pi/2+m\pi$, where $m$ is an integer, we have $c_1 =0$ and $f_{\mathrm{vp}}=0$ in Eq.~\ref{free energy}. Besides, the $c'_1$ term is small when $\frac{\Delta_{\mathrm{sc}}}{E_A} $ which can be achieved when $E_A$ is large and the intravalley inversion symmetry breaking is weak. Therefore, $f_{\mathrm{as}} \approx 0$. If both of these two conditions are satisfied, the second harmonic term dominates $F(\phi)$ and a $\pi$-periodic Josephson junction can be realized. In [Fig.~\ref{Fig.1D}(a)], the temperature dependence of the energy-phase relation of a 1D Josephson junction in the $c_1 \approx c'_1 \approx 0$ regime is depicted. The $\pi$-periodicity can be clearly seen and the energy-phase relation resembles the schematic illustration of [Fig.~\ref{fig:fig1}(d)]. At low temperatures, the higher Matsubara frequencies are important but the $\pi$-periodicity of the energy-phase relations is essentially the same as in the high temperature cases. On the other hand, if the intravalley inversion symmetry breaking is large ($E_A$ is small), the $ c'_1$ term can be significant. In this case, the energy-phase relation for different $E_A$ is shown in Fig.~\ref{Fig.1D}(b). This case corresponds to the situation of Fig.~\ref{fig:fig1}(e).  

\emph{$0$-$\pi$ Qubit.}--- In this section, we construct the $0$-$\pi$ qubit using the valley-polarized Josephson junction. As shown in Fig.~\ref{fig:fig1}(a), the qubit consists of a valley-polarized Josephson junction shunted by a capacitor. The Hamiltonian of the qubit is 
\begin{equation}
\label{qubit H}
    \hat{H}_{n_g}=4E_C (\hat{n}-n_g)^2+F(\hat{\phi})
\end{equation}
where $\hat{n}$ is the charge number operator, and $n_g$ is the offset charge of the Josephson junction. $E_C=\frac{e^2}{2C}$ is the charging energy and $C=C_J+C_S$ is the total capacitance with $C_J$ and $C_S$ the capacitance of the Josephson junction and the shunted capacitor, respectively. The wave function $\psi(\phi)$ of qubit satisfies the boundary condition $\psi(\phi)=\psi(\phi+2\pi)$. By using the transform $\psi=e^{in_g\phi}\Tilde{\psi}$, we can eliminate $n_g$ in Eq.~(\ref{qubit H}) and the boundary condition is $\Tilde{\psi}(\phi+2\pi)=e^{-i2\pi n_g}\Tilde{\psi}(\phi)$, which has the same form as the equation describing a particle moving in phase space with a periodic potential. Hence the eigenstates form bands with $-n_g$ as the crystal momentum. The bandwidth of the band reflects the sensitivity of the qubit to charge noise. As the transmon case \cite{PhysRevA.76.042319}, we require the shunted capacitance to be large enough so that the bandwidth \cite{Supplementary} $\delta\epsilon/\epsilon\simeq\exp(-\int d\phi\sqrt{\big(F(\phi)-\epsilon\big)/4E_C})\ll 1$, where $\epsilon$ is the energy level of qubit in the absence of $n_g$. For conventional transmons, this condition is reduced to $E_J/E_C\gg 1$ due to the simple form of its energy-phase relation $F(\phi)=-E_J\cos(\phi)$. Under this condition, the qubit is robust to charge noise.

As discussed above, for a valley-polarized Josephson junction with $\frac{2\Delta_{\mathrm{vp}}}{E_T}\approx\frac{\pi}{2}$, $F(\hat{\phi})$ is $\pi$-periodic as shown in Fig.~\ref{Fig.1D}(a).  At high temperature the energy-phase relation is dominated by the $\cos(2\phi)$ term. As the temperature becomes lower, higher order terms also make contributions to $F_J(\phi)$ and $\cos(2m\phi)$-terms become significant but the $\pi$-periodicity reminds. When $\frac{2\Delta_{\mathrm{vp}}}{E_T}$ deviates from $\frac{\pi}{2}$, the degeneracy between the two local minima is broken and the resulting free energy is shown in Fig.~\ref{Fig.1D}(c). The energy and the wavefunctions for the qubit states
for gate-defined $0$-$\pi$ qubit in MATBG will be presented
at the end of this work.

\emph{Transmon-like Qubit.}--- If the intravalley inversion symmetry of the valley-polarized weak link is significantly broken so that $E_A\simeq\Delta_{\mathrm{sc}}$, leading to a $\sin(\phi)$ term comparable to the $\cos(2\phi)$ terms in the energy-phase relation. As shown in Fig.~\ref{Fig.1D}(b) there are still two local minima within a period which are degenerate when $\Delta_{\mathrm{vp}}/E_T=\pi/4$ , but the barrier at $\phi=\frac{3\pi}{2}$ becomes much lower than the barrier at $\phi=\frac{\pi}{2}$. Assuming there is no tunneling between the states at the two local minima of $F(\phi)$, the ground state and the first excited state energy are denoted as $E_0$ and $E_1$, respectively. With tunneling, the qubit states of the transmon qubit are the symmetric and anti-symmetric linear combinations of the two states with energy $E_0$. Therefore, the qubit states are well separated from the higher energy states with energy in the order of $E_1$ and result in a large anharmonicity as shown in the Supplementary Material.

%%%%%%%%%%%%%%%%%%%%%%%%%%%%%%%%%%%%%%%%%%%%%%

\begin{figure}[t]
% \centering 
\includegraphics[width=1\linewidth]{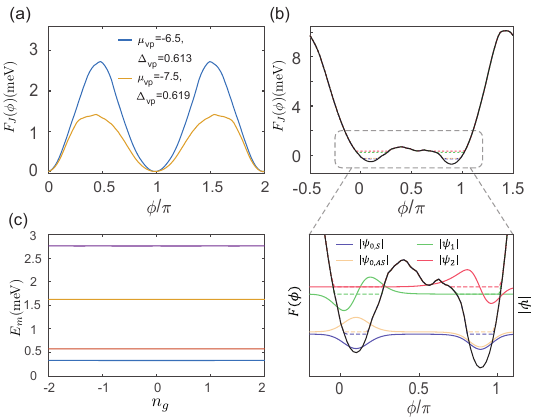}
    \caption{(a) Energy-phase relations in the $0$-$\pi$ qubit regime for two sets of parameters for the MATBG-based Josephson junction. The units for $\mu_{\mathrm{vp}}$ and $\Delta_{\mathrm{vp}}$ is meV. (b) Energy-phase relations for valley polarized Josephson junction in the transmon-like regime. Here, $\mu_{\mathrm{vp}}=-7.5$ meV, $E_C = 2\times10^3 $ MHz and $\Delta_{\mathrm{vp}}=0.363$ meV. The lower panel shows the wavefunction and the energy of the four lowest eigenstates of the corresponding qubit in the transmon-like regime. The energy levels of the four states are denoted by the horizontal dashed lines.  (c) The first four eigenenergies $E_m$ ($m=0,1,2,3$) as a function of offset charge $n_g $ of a valley polarized MATBG Josephson qubit with charging energy $E_C = 0.2\Delta_{sc}$. It is clear that the eigenenergies are insensitive to the change of $n_g $ such that the qubit is protected from change noise.
    }
    \label{Fig.3}
\end{figure}

%%%%%%%%%%%%%%%%%%%%%%%%%%%%%%%%%%%%%%%%%%%%%%%%

\emph{Energy-phase relation of gate-defined Josephson junction of MATBG}---In this section, we  demonstrate the feasibility of realizing the energy-phase relation discussed in the previous sections in realistic  gate-defined Josephson junctions of MATBG \cite{Jinxin_JJ_Diode,Yingming_phi_0_junction}.  The essential properties of moir\'{e} bands in MATBG can be captured by a tight binding model on a honeycomb lattice. The Hamiltonian of the model can be expressed as

\begin{equation}
    \begin{aligned}
        H_0 =& \left[ \sum_{\tau,\langle ij\rangle} t_{1} c^{\dagger}_{i\tau} c_{j\tau} + \sum_{\tau,\langle ij\rangle'} (t_{2} -i\tau t_2') c^{\dagger}_{i\tau} c_{j\tau} \right] \\
        &+ h.c. - \sum_{i,\tau} \mu c^{\dagger}_{i\tau} c_{i\tau}.
    \label{eq:tight_binding_H_main_text} 
    \end{aligned}
\end{equation}

Here, $\langle ij \rangle$ denotes the nearest hopping with amplitude $t_1$, and $\langle ij \rangle'$ sums over the fifth nearest hopping with amplitude $t_2-i\tau t_2'$\cite{Yuan_Fu:Wannier_Orbital, PhysRevB.98.045103}. The $\tau$ labels the two p-wave like orbital $p_x \pm i\tau p_y$, which is a representation of two valleys $\tau = \pm K$. The creation and annihilation operators of the electrons on site $i$ with valley index $\tau$ are denoted by $c^\dagger_{i\tau}$ and $c_{i\tau}$, respectively. In MATBG, the bandwidth is approximately 10 meV after considering lattice relaxation. To demonstrate the $0$-$\pi$ qubit regime, we choose $t_1 = 4$ meV, $t_2 = 0.02$ meV and $t_2'=0$. Note that $t_2'$ represents the warping strength of the Fermi surface, which is set to be zero in this case. In our model, the superconducting conducting region has a superconducting order parameter $\Delta_{\mathrm{sc}} $ and the weak link region has a  valley polarization order parameter $\Delta_{\mathrm{vp}}$. The energy-phase relation of the 2D model can be obtained using the lattice Green function method \cite{Supplementary}. 

We demonstrate that the energy-phase relation of the $0$-$\pi$ qubit is highly feasible in the MATBG, and the energy-phase relation with two different sets of parameters are shown in Fig.~\ref{Fig.3}~(a). The smallness of the trigonal warping term (controlled by $t_2'$) allows the realization of the $0$-$\pi$ qubit. This $0$-$\pi$ qubit is expected to be protected from decoherence originated from local perturbations \cite{kitaev2006protected,PhysRevA.87.052306,PhysRevB.90.094518,groszkowski2018coherence}. Moreover, since only fully gapped superconductors are involved, we expect the quasiparticle poisoning-induced decoherence can be suppressed exponentially when the operating temperature scale is much smaller than the pairing gap $\Delta_{\mathrm{sc}} $.

On the other hand, for large $t_2' = 0.8$ meV, $t_1 = 4$ meV and $t_2 = 0.2$ meV, the energy-phase relation is shown in Fig.~\ref{Fig.3}~(b). When the junction is shunted by a capacitor with $E_C = 2\times10^3 $ MHz, the wavefunctions of the qubit states are shown in Fig.~\ref{Fig.3}~(b). The qubit state wavefunctions $\Psi_{0S}$ and $\Psi_{0AS}$, which are the symmetric and the anti-symmetric wavefunctions with respect to $\phi$, are well separated from the excited states $\Psi_1$ and $\Psi_2$ in energy.  The energy splitting of the two-qubit states is $0.01E_C$ and the energy gap between the qubit states and the excited states is about $60 E_C$. The horizontal dashed lines in Fig.~\ref{Fig.3}~(b) indicate the energy levels of different states. These energy levels are extremely stable against the change of $n_g$ [Fig.~\ref{Fig.3}~(c)], resulting in long decoherence time against charge fluctuations \cite{Supplementary}.

\emph{Discussion.}--- Recently, qubit realizations using a single Josephson junction with unconventional energy-phase relation have attracted much attention \cite{PhysRevB.105.L180502, patel2023dmon, brosco2023superconducting}. In particular, at the concluding stage of this work, we came across proposals using d-wave superconductors to construct Josephson junctions with unconventional energy-phase relations \cite{patel2023dmon, brosco2023superconducting}. On the other hand, the proposed gate-defined MATBG-based qubits do not require external magnetic fields nor unconventional superconductors. Importantly, the proposed Josephson junction had been experimentally realized \cite{2023NatCo..14.2396D}. In addition, the gatemons, based on semiconductor nanowires \cite{PhysRevLett.115.127001, PhysRevApplied.18.034042, PhysRevB.108.L020505} are also  transmon-like qubits which can be manipulated through electrostatic gating. However, there is no unconventional energy-phase relation in gatemons and they are not related to the $0$-$\pi$ qubits discussed in this work.

\emph{\bf Acknowledgments.}
K.T.L. acknowledges the support of the Ministry of Science and Technology, China, and the Hong Kong Research
Grants Council through Grants No. 2020YFA0309600, No. RFS2021-6S03, No. C6025-19G, No. AoE/P-701/20, No. 16310520, No. 16310219 and No. 16307622.

\end{document}